\documentclass[12pt,oneside]{article}
\usepackage{amsfonts,amssymb,graphicx}
\def\be{\begin{equation}}
\def\ee{\end{equation}}
\def\bea{\begin{eqnarray}}
\def\eea{\end{eqnarray}}
\def\c{\bar{c}}
\def\<{\langle}
\def\>{\rangle}
\def\~{\tilde}
\def\s{\sigma}

\def\a{\alpha}
\def\b{\beta}

\def\o{\omega}
\def\t{\tau}

\newcommand{\qed}{\hfill \ensuremath{\Box}}

\newcommand{\Z}{\Bbb Z}

\newcommand{\av}[1]{\mbox{{\rm Av}}\left(#1\right)}

\newtheorem{remark}{Remark}

\newtheorem{theorem}{Theorem}

\newtheorem{lemma}{Lemma}

\begin{document}
\begin{center}
\vspace{1truecm}
{\bf\sc\Large the ghirlanda-guerra identities}\\
\vspace{1cm}
{Pierluigi Contucci$^{\dagger}$, Cristian Giardin\`a$^{\ddagger}$}\\
\vspace{.5cm}
{\small $\dagger$ Dipartimento di Matematica} \\
    {\small Universit\`a di Bologna,
    40127 Bologna, Italy}\\
    {\small {e-mail: {\em contucci@dm.unibo.it}}}\\
\vspace{.5cm}
{\small $\ddagger$ EURANDOM}\\
{\small P.O. Box 513 - 5600 MB Eindhoven, The Netherland}\\
{\small {e-mail: {\em giardina@eurandom.tue.nl}}}
\vskip 1truecm
\end{center}
\vskip 1truecm
\begin{abstract}\noindent
If the variance of a Gaussian spin-glass Hamiltonian grows like the volume
the model fulfills the Ghirlanda-Guerra identities in terms of the
normalized Hamiltonian covariance.
\end{abstract}
\newpage\noindent
\section{Introduction}
In the last decade new ideas and technical methods have been developed in
the attempt to build the spin glass theory on rigorous
mathematical grounds. The most recent example is the interpolation strategy
\cite{GT} introduced to prove the
existence of the thermodynamic limit in the Sherrington-Kirkpatrick \cite{SK}
model
and its use in the proof of
the Parisi free energy \cite{MPV} of the same model \cite{G1,T1} and in the
identification of an extended variational principle \cite{AiSS}.

The first basic contribution in the field came with the work by
Guerra \cite{G2} on how to prove some correlation identities of the
Sherrington-Kirkpatrick model that were only assumed within the
ultrametric structure of the Parisi solution. Those identities were later
generalized by Ghirlanda-Guerra \cite{GG} and
are playing an increasingly important role in the mathematical approach to
the low temperature spin glass phase (see \cite{B,T2} and references therein).
The Ghirlanda-Guerra identities are consequence of a very basic principle of
statistical mechanics i.e. the vanishing of the
fluctuation of the energy per particle: at increasing volumes the energy per
particle approaches a constant with respect to the
equilibrium measure. Within the non-disordered classical cases all that
simply implies the
finiteness of the specific heat almost everywhere in
the temperature (see nevertheless the implications in classical mean-field
models \cite{CGI}); however in the spin glass cases, where the
equilibrium {\it quenched} state is a properly intertwined composition of
the Boltzmann-Gibbs and the disorder measures,
its consequences are way more subtle. The work \cite{G2} led to the
identification of the {\it stochastic stability}
\cite{AC}, an invariance property of the quenched state under a class of suitable perturbations. 
Subsequently stochastic stability was developed and used clarify the relation between the equilibrium
and the off-equilibrium properties in the spin glass
phase \cite{FMPP1,FMPP2}. Recently stochastic stability has been classified
\cite{C,BCK} and placed on rigorous grounds in \cite{CGi}, where 
its relation with the mentioned identities is also discussed.

In this work we obtain a condition that guarantee the validity of the
Ghirlanda-Guerra identities: our result states that they hold true whenever
the variance of the Hamiltonian function grows like the volume.
Such a condition is the same that ensure the existence (boundedness) of the
thermodynamic limit \cite{CG} and applies to every spin glass
model studied so far: to the Edwards-Anderson
model, to the finite dimensional cases with summable or non-summable
interactions in the sense of Kanin and Sinai \cite{KS}, to the mean field
cases like SK, p-spin, REM and GREM,
up to the general spin glass  model of subset interaction on which our
general condition has been tailored. It is important to
stress that the identities we prove hold in terms of the normalized 
Hamiltonian covariance which has a different spin expression in each model:
for instance in the SK model it coincides with the square power of the overlap function,
while for the EA model it is the link overlap \cite{C2,NS}. The strategy we
use to achieve the result relies on very simple methods like the bound on martingale sums
and classical inequalities.

The paper is organized in definitions (Sec. 2), results (Sec. 3), proofs
(Sec. 4) and is concluded with some comments
and perspectives (Sec. 5).

\section{Definitions}

We consider a disordered model of Ising configurations
$\s_n=\pm 1$, $n\in \Lambda\subset \Z^d$ for some $d$-parallelepiped
$\Lambda$ of volume $|\Lambda|$. We denote
$\Sigma_\Lambda$ the set of all $\s=\{\s_n\}_{n\in \Lambda}$, and
$|\Sigma_\Lambda|=2^{|\Lambda|}$. In the sequel the
following definitions will be used.

\begin{enumerate}

\item {\it Hamiltonian}.\\ For every $\Lambda\subset \Z^d$ let
$\{H_\Lambda(\sigma)\}_{\s\in\Sigma_N}$
be a family of
$2^{|\Lambda|}$ {\em translation invariant (in distribution) centered
Gaussian} random variables defined, in analogy with \cite{RU}, according to
the very general representation
\be
H_{\Lambda}(\s) \; = \; - \sum_{X\subset \Lambda} J_X\s_X
\label{hami}
\ee
where 
\be
\s_X=\prod_{i\, \in X}\s_i \; ,
\ee
($\s_\emptyset=0$) and the $J$'s are independent Gaussian variables with
zero mean
\be
{\rm Av}(J_X) = 0 \; ,
\ee 
and (translation invariant) variance
\be
{\rm Av}(J^2_X) = \Delta^2_X  \; .
\ee
\item {\it Covariance matrix}.
\be\label{cc}
{\cal C}_\Lambda (\s,\tau) \; := \; \av{H_\Lambda(\s) H_\Lambda (\tau)}  \;
= \; \sum_{X\subset\Lambda}\Delta^2_X\s_X\t_X\, .
\ee
By the Schwartz inequality
\be\label{sw}
|{\cal C}_\Lambda (\s,\t)| \; \le \; \sqrt{{\cal C}_\Lambda
(\s,\s)}\sqrt{{\cal C}_\Lambda (\t,\t)} \; = \;
\sum_{X\subset\Lambda}\Delta^2_X
\ee 
for all $\s$ and
$\t$.
\item {\it Thermodynamic Stability}.\\
The Hamiltonian (\ref{hami}) is thermodynamically stable if it exist a
constant $\bar{c} < \infty $ such that
\be\label{thst}
\sup_{\Lambda\subset\Z^d} \frac{1}{|\Lambda|} {\cal C}_\Lambda (\s,\s) \; =
\; 
\sup_{\Lambda\subset\Z^d}
\frac{1}{|\Lambda|}\sum_{X\subset\Lambda}\Delta^2_X
\; \le \; \c < \infty
\ee
Together with translation invariance a condition like the (\ref{thst}) is
equivalent to
\be
\sum_{X\ni 0}\frac{\Delta^2_X}{|X|} \; \le \; \c \; .
\ee
In fact
\be
\sum_{X\subset\Lambda}\Delta^2_X \; = \; \sum_{x\in\Lambda}\sum_{X\ni
x}\frac{\Delta^2_X}{|X|} \; = \; |\Lambda|\sum_{X\ni
0}\frac{\Delta^2_X}{|X|} \; .
\ee
Alternatively, summing over the equivalence classes $\widetilde X$ of the
translation group, the (\ref{thst}) is equivalent to
\be
\label{equiv}
\sum_{\widetilde X}\Delta^2_{\widetilde X} \; \le \; \c \; .
\ee
Thanks to the (\ref{sw}) a thermodynamically stable model fulfills the bound
\be
{\cal C}_\Lambda (\s,\t) \; \le \; \c \, |\Lambda|
\ee
and has a order $1$ normalized covariance
\be
c_{\Lambda}(\s,\t) \; : = \; \frac{1}{|\Lambda|}{\cal C}_\Lambda (\s,\t)
\ee

\item {\it Random partition function}.
\be\label{rpf}
{\cal Z}(\beta) \; := \; \sum_{\s  \in \,\Sigma_\Lambda}
e^{-\beta{H}_\Lambda(\s)}
\; .
\ee
\item {\it Random free energy}.
\be\label{rfe}
-\beta {\cal F}(\beta) \; := \; {\cal A}(\beta) \; := \; \ln {\cal Z}(\beta)
\; .
\ee
\item {\it Random internal energy}.
\be\label{rie}
{\cal U}(\beta) \; := \; \frac{\sum_{\s  \in \,\Sigma_\Lambda}
H_{\Lambda}(\s)e^{-\beta{H}_\Lambda(\s)}}{\sum_{\s  \in \,\Sigma_\Lambda}
e^{-\beta{H}_\Lambda(\s)}}
\; .
\ee
\item {\it Quenched free energy}.
\be
-\beta F(\beta) \; := \; A(\beta) \; := \; \av{ {\cal A}(\beta) }\; .
\ee
\item $R$-{\it product random Gibbs-Boltzmann state}.
\be
\Omega (-) \; := \;
\sum_{\sigma^{(1)},...,\sigma^{(R)}}(-)\,
\frac{
e^{-\beta[H_\Lambda(\s^{(1)})+\cdots
+H_\Lambda(\sigma^{(R)})]}}{[{\cal Z}(\beta)]^R}
\; .
\label{omega}
\ee
\item {\it Quenched equilibrium state}.
\be
<-> \, := \av{\Omega (-)} \; .
\ee
\item\label{obs} {\it Observables}.\\
For any smooth bounded function $G(c_{\Lambda})$
(without loss of generality we consider $|G|\le 1$ and no assumption of
permutation invariance
on $G$ is made) of the covariance matrix entries we introduce the random
(with respect to
$<->$) $R\times R$ matrix of elements $\{q_{k,l}\}$ (called {\it generalized
overlap})
by the formula
\be
<G(q)> \; := \; \av{\Omega (G(c_{\Lambda}))} \; .
\ee
E.g.: 
$G(c_\Lambda)=c_{\Lambda}(\sigma^{(1)},\sigma^{(2)})c_{\Lambda}(\sigma^{(2)}
,\sigma^{(3)})$
\be
<q_{1,2}q_{2,3}> \; = \;
\av{\frac{\sum_{\sigma^{(1)},\sigma^{(2)},\sigma^{(3)}}
c_{\Lambda}(\sigma^{(1)},\sigma^{(2)})c_{\Lambda}(\sigma^{(2)},\sigma^{(3)})
\,
e^{-\beta[\sum_{i=1}^{3}H_\Lambda(\
\s^{(i)})]}}{[{\cal Z}(\beta)]^3}}
\ee
\end{enumerate}

\section{Results}
The Ghirlanda-Guerra identities admit several equivalent formulations. They
can be expressed in terms
of factorization properties of the quenched distribution of the generalized
overlap \cite{GG,B,T2} as well as in terms
of expectations of observables. In this work we chose the second approach
because it allows to distinguish the identities in two classes with
different physical meaning: the first expresses the regularity 
with respect to the temperature, the second the self-averaging of intensive
quantities.\\

In relation to the definitions of the previous section
it holds the following:
\begin{theorem}
The quenched equilibrium state of a thermodynamically stable Hamiltonian
fulfills, for every observable $G$ and
every temperature interval $[\b_1^2,\b_2^2]$ the following identities in the
thermodynamic limit
\be\label{st1}
\int_{\b_1^2}^{\b_2^2}<\mathop{\sum_{k,l=1}^{R}}_{k\ne l} G \,q_{\,l,\,k}
- 2R G \, \sum_{l=1}^{R} q_{\,l,\,R+1}
+R(R+1) G \, q_{\,R+1,\,R+2} > d\b^2\; = \; 0
\ee
\be\label{st2}
\int_{\b_1^2}^{\b_2^2}
 \left [
\sum_{k=1}^R  <G \, q_{\,k,\,R+1}> - (R+1)  <G \, q_{\,R+1,\,R+2}>
 +  <G>\,<q_{1,2}> 
\right] d\b^2 \; = \; 0
\ee
\end{theorem}
\vspace{.5truecm}
\begin{remark}
The two previous relations when applied to $G(q)=q_{1,2}$ combined together
lead to the well known \cite{MPV,G2}:
\be
<q_{1,2}q_{2,3}> \; = \; \frac{1}{2}<q^2>+\frac{1}{2}<q>^2
\ee
\be
<q_{1,2}q_{3,4}> \; = \; \frac{1}{3}<q^2>+\frac{2}{3}<q>^2
\ee
\end{remark}
\begin{remark}
It is straightforward to verify that the condition (\ref{thst}) of
thermodynamic stability holds for all the known
spin glass models. Here are a few examples:
\begin{enumerate}
\item The Edwards-Anderson model \cite{EA}. The nearest neighbor case
is defined by $\Delta^2_X=1$ if $X=(n,n')$ and $|n-n'|=1$. The condition
(\ref{thst}) is verified by $\c=d$.
\item More generally one consider still a two body interaction with
$\Delta^2_X=|n-n'|^{-2d\a}$. The regime $\a>1$ comes from a summable
interaction. The condition (\ref{thst})
is verified by $\c=(2\a-1)^{-d}$ for all $\a>1/2$ thus including also the
non summable case \cite{KS}.
\item The SK model \cite{SK}. Although it is not a finite dimensional model
it may
still be embedded in $\Z$,
with $\Delta^2_X=0$ unless $|X|=2$ and $\Delta_{i,j}=N^{-1}$ with $N=|\Lambda|$. It obviously
fulfills condition (\ref{thst}) with $\c=1$.
\item The p-spin. Analogously as above $\Delta_X=0$ if $|X|\neq p$ and
$\Delta^2_X=1/N^p$ otherwise. It is thermodynamically stable with $\c=1$
\item The REM \cite{D} and GREM \cite{DG} models. Although they have not
been defined as spin
models their discrete nature allows
to associate to them a spin Hamiltonian.
For instance it is easy to prove that the REM is represented by
by the Hamiltonian (\ref{hami}) with $\Delta_X^2 = N\, 2^{-N}$
which satisfies the condition (\ref{thst}) with $\c=1$, see also \cite{B}.
The same argument holds for the GREM  \cite{CDGG,CG2} 
which is again thermodynamically stable with $\c=1$.
\end{enumerate}
\end{remark}
\begin{remark}
The relevance of the identities is evident considering that they
reduce the degrees of freedom a priori carried by each spin glass model. 
In the mean field case for instance the method led
to the rigorous proof of a property called {\it replica equivalence}
\cite{Pa,C2} which can viewed as an ansatz generalizing the ultrametric one. 
The purely ultrametric identities (still lacking
a rigorous mathematical derivation) which are built in the Parisi solution
of the SK model are not contained in the Ghirlanda-Guerra ones.
\end{remark}
\begin{remark}
It would be interesting to establish, or disprove, the same identities in a 
stronger sense, i.e. everywhere in the temperature. One of the limit of the 
method used to achieve our results is that it is intrinsically restricted to 
hold in $\b$-average, i.e. in every interval excluding
at most isolated singularities. It is still an open question if, in
the spin glass phase, there are similar singularities or if the identities 
hold just everywhere. The only existing
results are evidences of numerical nature of the validity of those
identities everywhere \cite{MPRRZ,CGi2}.
\end{remark}
\section{Proof}
The statements (\ref{st1}) and (\ref{st2}) are proved respectively in the
lemmata of
subsections \ref{ss1} and \ref{ss2}. The proof uses only elementary methods
like martingale differences and
classical inequalities. Let
$h(\s)=|\Lambda|^{-1}H_\Lambda(\s)$ be the Hamiltonian per particle. We
consider the quantity
\be
\sum_{l=1}^R \left\{<h(\s^{(l)}) \; G> - <h(\s^{(l)})><G>\right\} =
\Delta_1 G +
\Delta_2 G
\ee
where
\be
\label{delta1}
\Delta_1 G = \sum_{l=1}^R \left\{\av{\Omega[h(\s^{(l)})\,G] -
\Omega[h(\s^{(l)})]\Omega[G])} \right\}
\ee
\be
\label{delta2}
\Delta_2 G = \sum_{l=1}^R \left\{\av{\Omega[h(\s^{(l)})]\Omega[G]} -
\av{\Omega[h(\s^{(l)})]} \av{\Omega[G]}\right\}
\ee
%
%
%
%

\subsection{Stochastic Stability Bounds, vanishing of $\Delta_1
G$}\label{ss1}
We follow the method of stochastic stability as developed in \cite{CGi}.
\begin{lemma}\label{l1}
For every bounded observable $G$, see definition (\ref{obs}), we have that
for every interval $[\beta_1,\beta_2]$ in the
thermodynamic limit
\be\label{ss}
\int_{\beta_1}^{\beta_2} \Delta_1 G \;d\beta \; = \; 0
\ee
\end{lemma}
{\bf Proof.} We observe that deriving $<G>$ with respect to the temperature
\be
- \frac{\partial <G>}{\partial \beta} \; = \; |\Lambda|
\sum_{l=1}^R \left\{\av{\Omega[h(\s^{(l)})\,G] -
\Omega[h(\s^{(l)})]\Omega[G])} \right\}
\ee
Integrating in $d\b$ 
we obtain thanks to (\ref{delta1})
\be
\int_{\beta_1}^{\beta_2} \Delta_1 G \; d\beta =
\frac{<G>(\beta_2)\; - <G>(\beta_1)}{|\Lambda|}
\ee
Remembering the assumption on boundedness of function $G$
this proves the lemma. \qed
\begin{remark}
The previous lemma is related to a general property of
disordered systems which is known as stochastic stability
(see \cite{AC,CGi}). It says that the equilibrium state in a spin glass
model is invariant under a suitable class of perturbation in all
temperature intervals of continuity.
\end{remark}
\vspace{0.3cm}
\noindent
\begin{lemma}\label{l2} The following expression holds:
\be\label{expr-delta1}
 \Delta_1 G \; =\; 
-\b < G\,\left[ \mathop{\sum_{k,l=1}^{R}}_{k\ne l}  \,q_{\,l,\,k}
- 2R  \, \sum_{l=1}^{R} q_{\,l,\,R+1}
+R(R+1) \, q_{\,R+1,\,R+2} \right] > \; .
\ee
\end{lemma}
{\bf Proof.}\\
For each replica $l$ $(1\le l\le R)$, we evaluate separately
the two terms in the right side of Eq. (\ref{delta1}) by using
the integration by parts (generalized Wick formula) for
correlated Gaussian random variables, $x_1,x_2,\ldots,x_n$
\be\label{ibp}
\av { x_i\, \psi(x_1,...,x_n) } =
\sum_{j=1}^n \av { x_i x_j  } \,
\av {\frac{\partial \psi(x_1,...,x_n)}{\partial x_j} } \;.
\ee
It is convenient to denote by $p\,(R)$ the Gibbs-Boltzmann weight
of R copies of the deformed system
\be
p\,(R) \,= \,
\frac{
e^{-\beta\,[\,\sum_{k=1}^R H_\Lambda(\s^{(k)})\,]}}
{[{\cal Z}(\beta)]^R} \;,
\ee
so that we have
\be
\label{derivataBolt}
- \frac{1}{\beta}\frac{dp\,(R)}{dH_{\Lambda}(\tau)} \;=\;
p\,(R) 
\left(\sum_{k=1}^R \delta_{\s^{(k)},\,\tau}\right)
- R \;p\,(R)\;
\frac{
e^{-\beta[H_\Lambda(\tau)]}}
{[{\cal Z}(\beta)]} \;.
\ee
We obtain
\bea
\av{\Omega(h(\s^{(l)})\,G)} & = &
\frac{1}{|\Lambda|}\,
\av{\;
\sum_{\sigma^{(1)},...,\sigma^{(r)}}\;
G\;H_{\Lambda}(\s^{(l)}) \;
p\,(R)} \\
& = & 
\label{line1}
\av{ \;
\sum_{\sigma^{(1)},...,\sigma^{(r)}}\;\sum_{\tau}\;G\;
{c}_{\Lambda}(\s^{(l)},\tau)\;
\frac{dp\,(R)}{dH_{\Lambda}(\tau)}}
\qquad\qquad
\\
& = & 
\label{line2}
- \beta \, \left[
\sum_{k=1}^{R} <G \,q_{\,l,\,k}> -
R  <G \, q_{\, l ,\, R+1}>
\right]
\eea
where in (\ref{line1}) we made use of the integration
by parts formula and (\ref{line2}) is obtained
by (\ref{derivataBolt}).
Analogously, the other term reads
\bea
\;
\av{\Omega(h(\s^{(l)}))\,\Omega (G)} & = &
\frac{1}{|\Lambda|}\,
\av{\;
\sum_{\sigma^{(l)}}\sum_{\tau^{(1)},...,\tau^{(R)}}\;
G\;H_{\Lambda}(\s^{(l)}) \;
p\,(R+1)} \\
& = & 
\label{line3}
\av{\;
\sum_{\sigma^{(l)}}\sum_{\tau^{(1)},...,\tau^{(R)}}\;\sum_{\gamma}\;G\;
{c}_{\Lambda}(\s^{(l)},\gamma)\;
\frac{dp\,(R+1)}{dH_{\Lambda}(\gamma)}}\quad
\qquad
\\
& = & 
\label{line4}
- \beta\,\left [
\sum_{k=1}^{R+1} <G \, q_{\,k,\,R+1}>
- (R+1) <G \, q_{\,R+1,\,R+2}>
\right]\nonumber\\
\eea
Inserting the (\ref{line2}) and (\ref{line4}) in Eq. (\ref{delta1})
we finally obtain the expression (\ref{expr-delta1}).
\qed
\subsection{Martingale Bounds, vanishing of $\Delta_2 G$}\label{ss2}
The method of the martingale differences to prove the self averaging of the
free energy, or in general to bound the fluctuations of extensive quantity,
has been
applied in the context of spin glasses in \cite{PS} for the SK case and in
\cite{WA}
in the case of finite dimensional models. Our formulation applies to both
cases and extends
the previous results. For instance our method includes the non summable
interactions in finite
dimensions \cite{KS} and the p-spin mean field model as well as the REM
\cite{D} and GREM \cite{DG} models.
\begin{lemma}\label{martin}
The free energy is a self averaging quantity, i.e. it exist a positive
function $c(\beta)$
such that
\be\label{sav}
V({\cal A}) \; = \;  \av{{\cal A}^2}-\av{{\cal A}}^2\le c(\b) {|\Lambda|}
\ee
\end{lemma}
{\bf Proof.} For an assigned volume $\Lambda$ we enumerate by the index $k$
the interacting subsets $X$ from
$1$ to $N_\Lambda$ and considering the random partition function (\ref{rpf})
we define
\be\label{akappa}
A_k \; = \; {\rm Av}_{\le k}\ln{\cal Z}(\beta) \; ,
\ee
where the symbol ${\rm Av}_{\le k}$ denotes the Gaussian integration
performed
only on the first $k$ random variables $J_X$.
Clearly $A_0={\cal A}(\b)$ and $A_{N_\Lambda}=A(\b)$. Introducing the
quantity
\be
\Psi_k \; = \; A_k-A_{k+1} \; ,
\ee
it holds 
\be
{\cal A}-\av{{\cal A}} \; = \; \sum_{k=0}^{N_\Lambda -1} \Psi_k
\ee
and
\be
V({\cal A}) \; = \; \sum_{k}\av{\Psi^2_k} \, + \, 2 \sum_{k >
k'}\av{\Psi_k\Psi_{k'}} \; .
\ee
First we observe that the second sum is zero, being zero each of its terms.
In fact
\be
\av{\Psi_k\Psi_{k'}}=\av{{\rm Av}_{\le k}(\Psi_k\Psi_{k'})}=\av{\Psi_k{\rm
Av}_{\le k}(\Psi_{k'})}
\ee
and
\be
{\rm Av}_{\le k}(\Psi_{k'}) \; = \; {\rm Av}_{\le k}(A_{k'}-A_{k'+1}) \; =
\; 0
\ee
thanks to the property
\be
{\rm Av}_{\le k}(A_{k'}) \; = \; A_k \qquad \forall \qquad k\ge k' \; .
\ee
We introduce now the interpolated Hamiltonian
\be
H^{(t)}_{\Lambda}(\s) \; = \; - \sum_{l=1}^{N_\Lambda} J_lt_{l}\s_l
\ee
with
\be
t_l \; = \; \left\{
\begin{array}{ll} 
t, & \mbox{if $l=k+1$}, \\
 1, & \mbox{otherwise} \, ,
\end{array}\right. 
\ee  
and define the quantity
\be
A_{k}(t) \; = \; {\rm Av}_{\le k}\ln\sum_{\s  \in \,\Sigma_\Lambda}
e^{-\beta H^{(t)}_{\Lambda}(\s) }
\; .
\ee
By the fundamental theorem of calculus
\be
A_{k} \; = \; A_{k}(0) + B_{k}
\ee
with
\be
B_{k} \; = \; \int_{0}^{1}\frac{dA_{k}(t)}{dt}dt \; = \;
\beta\int_{0}^{1}{\rm Av}_{\le k}\,\o_{t}(J_{k+1}\s_{k+1}) \; .
\ee
We observe
\bea
\av{\Psi^2_k} \; = \; \av{[A_k-A_{k+1}]^2} \; = \; \av{{\rm
Av}_{k+1}[A_k-{\rm Av}_{k+1}A_k]^2}\\\nonumber
\; = \; \av{{\rm Av}_{k+1}(A^2_k)-[{\rm Av}_{k+1}(A_k)]^2}
\eea
Since $A_{k}$ and $B_k$ differ by a constant with respect to ${\rm
Av}_{k+1}$ (integration with respect to the $(k+1)$-th
Gaussian)
we have that their variance is the same:
\be
\av{\Psi^2_k} \; = \; \av{{\rm Av}_{k+1}(A^2_k)-[{\rm Av}_{k+1}(A_k)]^2} \; = \; \av{{\rm
Av}_{k+1}(B^2_k)-[{\rm Av}_{k+1}(B_k)]^2} \; .
\ee
We will estimate separately the two terms $\av{{\rm Av}_{k+1}(B^2_k)}$ and
$\av{[{\rm Av}_{k+1}(B_k)]^2}$.
By a simple integration by parts (\ref{ibp}) on $J_{k+1}$ we obtain
\be
{\rm Av}_{k+1}(B_k) \; = \; \beta^2\Delta^2_{k+1}\int_{0}^{1}{\rm Av}_{\le
k+1}\,[1-\o^2_{t}(\s_{k+1})]tdt \; \le \;
\frac{1}{2}\b^2\Delta^2_{k+1}
\ee
which implies
\be\label{aqb}
0 \; \le \; \av{[{\rm Av}_{k+1}(B_k)]^2} \; \le \; \frac{1}{4}\b^4\Delta^4_{k+1} \; .
\ee
Analogously we have
\be
{\rm Av}_{k+1}(B^2_k) \; = \; {\rm Av}_{k+1}\int_{0}^{1}\int_{0}^{1}\,{\rm
Av}_{\le k}(\o_t(J_{k+1}\s_{k+1}))
{\rm Av}_{\le k}(\o_s(J_{k+1}\s_{k+1}))stdsdt
\ee
Applying twice the integration by parts (\ref{ibp}) we get
\bea\label{abq}
{\rm Av}_{k+1}(B^2_k) \; &=& \; \b^2\Delta^2_{k+1}{\rm
Av}_{k+1}\int_{0}^{1}\int_{0}^{1}\,{\rm Av}_{\le k}(\o_t(\s_{k+1}))
{\rm Av}_{\le k}(\o_s(\s_{k+1}))stdsdt +\\\nonumber
&-& 2 \b^4\Delta^4_{k+1}{\rm Av}_{k+1}\int_{0}^{1}\int_{0}^{1}\,{\rm
Av}_{\le k}(\o_t(\s_{k+1})[1-\o^2_t(\s_{k+1})])
{\rm Av}_{\le k}(\o_s(\s_{k+1}))st^3dsdt +\\\nonumber
&-& 2 \b^4\Delta^4_{k+1}{\rm Av}_{k+1}\int_{0}^{1}\int_{0}^{1}\,{\rm
Av}_{\le k}(\o_s(\s_{k+1})[1-\o^2_s(\s_{k+1})])
{\rm Av}_{\le k}(\o_t(\s_{k+1}))s^3tdsdt +\\\nonumber
&+& 2 \b^4\Delta^4_{k+1}{\rm Av}_{k+1}\int_{0}^{1}\int_{0}^{1}\,{\rm
Av}_{\le k}[1-\o^2_s(\s_{k+1})]{\rm Av}_{\le
k}[1-\o^2_s(\s_{k+1})]s^2t^2dsdt \\\nonumber
&\le & \frac{1}{4}\b^2\Delta^2_{k+1}+\frac{13}{18}\b^4\Delta^4_{k+1}
\eea
Putting together the (\ref{aqb}) and the (\ref{abq}) we find
\be
\av{\Psi^2_k} \; \le \; \frac{1}{4}\b^2\Delta^2_{k+1}+\frac{35}{36}\b^4\Delta^4_{k+1}
\ee
\be
V({\cal A}) \; = \; \sum_{k}\av{\Psi^2_k} \; \le \; \sum_{X\subset \Lambda }
\frac{1}{4}\b^2\Delta^2_{X}+\frac{35}{36}\b^4\Delta^4_{X} \; .
\ee
By the assumption of thermodynamic stability with the formulation (\ref{equiv})
and using the inequality 
$\sum_{\widetilde X}\Delta^4_{\widetilde X} \le 
(\sum_{\widetilde X}\Delta^2_{\widetilde X})^2$
we obtain
\be
V({\cal A}) \; \le \; |\Lambda|(\frac{1}{4}\b^2 \c +\frac{35}{36}\b^4\c^2)
\ee
which fulfills (\ref{sav}) with $c(\b)=\frac{1}{4}\b^2 \c +\frac{35}{36}\b^4\c^2$. \qed
\begin{lemma}\label{sa}
The internal energy is self averaging almost everywhere $\b$, i.e. defining
$u={\cal U}/|\Lambda|$ and
$V(u)=\av{u^2}-\av{u}^2$ it holds in the thermodynamic limit
\be\label{vtoz}
\int_{\beta_1}^{\beta_2} V(u) \, d\b \; \to \; 0
\ee
\end{lemma}
{\bf Proof.}\\ 
The result is obtained in two steps which use general theorems of measure
theory. First from lemma \ref{martin} we obtain the
convergence to zero almost everywhere (in $\b$) of the variance of the
internal energy, then thanks to a bound on the
variance of the internal energy we apply the Lebesgue dominated convergence
theorem which gives the lemma statement.
The sequence of convex functions ${\cal A}(\beta)/|\Lambda|$ converges a.e.
(in $J$) to the limiting value $a(\beta)$ of its
average
\cite{GT,CG} and the convergence is self averaging in the sense of lemma
\ref{martin}. By general convexity arguments \cite{RU}
it follows that the sequence of the derivatives ${\cal A}'(\beta)/|\Lambda|$
converges to $u(\beta)=a'(\beta)$ almost everywhere
in $\beta$ and also that the convergence is self averaging. In fact the
vanishing of the variance of a sequence
of convex functions is inherited, in all points in which the derivative
exists (which is almost everywhere for a convex function),
to the sequence of its derivatives (see \cite{S,OTW}). From lemma
\ref{martin} we have then
\be
V(u) \; \to \; 0 \quad \beta \;-\;  a.e.
\ee 
In order to obtain the convergence in $\beta$-average we use the Lebesgue
dominated convergence theorem. In fact
we prove that the sequence of variances of $u$ is uniformly bounded (in
every interval $[\b_1,\b_2]$) by an integrable
function of $\beta$.
A lengthy but simple computation which uses again integration by parts gives
\be\label{u}
\av{{\cal U}} \; = \; \av{\sum_{X\subset \Lambda} J_X\o(\s_X)} \; = \;
\sum_{X\subset\Lambda}\b\Delta_X^2[1-\av{\o^2(\s_X)}] \; \le
\b|\Lambda|\c
\ee
\bea\label{u2}
\av{{\cal U}^2} \; &=& \; \av{\sum_{X,Y\subset \Lambda}
J_XJ_{Y}\o(\s_X)\o(\s_Y)} \; = \; \\\nonumber
\; &=& \; 
\sum_{X,Y\subset\Lambda}\b^2\Delta_X^2\Delta_Y^2 {\rm
Av}\left[1-\o^2(\s_X)-\o^2(\s_Y)+6\o^2(\s_X)\o^2(\s_Y)+\right.\\
&-&\left. 6 \o(\s_X)\o(\s_Y)\o(\s_X\s_Y)+\o^2(\s_X\s_Y)\right] \; \le \; 14
\b^2|\Lambda|^2 \c^2
\eea
from which
\be
V(u) \; \le \; 15\b^2\c^2 \; .
\ee
From this follows (\ref{vtoz}). \qed
\begin{lemma}\label{d2}
For every bounded observable $G$, see definition (\ref{obs}), we have
that for every interval $[\beta_1,\beta_2]$ in the thermodynamic
limit
\be
\int_{\beta_1}^{\beta_2}\Delta_2 G \, d\beta \; = \; 0
\ee
\end{lemma}
{\bf Proof}.\\ 
Thanks to the Schwartz inequality
\bea
\Delta_2 G \; &=& \; \av{uG-\av{u}\av{G}} \; = \;
\av{\left[u-\av{u}\right]\left[G-\av{G}\right]} = \\\nonumber
&\le& \sqrt{\av{\left[u-\av{u}\right]^2}}\sqrt{\av{\left[G-\av{G}\right]^2}}
\; \le \; \sqrt{2}\sqrt{V(u)}
\eea
\be
(\Delta_2 G)^2 \; \le \; 2 V(u)
\ee
\be
\left| \int_{\beta_1}^{\beta_2} \Delta_2 G d\b\right| \; \le
\sqrt{\int_{\beta_1}^{\beta_2} (\Delta_2 G)^2d\b}\sqrt{\b_2-\b_1} \; \le \;
\sqrt{2(\b_2-\b_1)} \sqrt{\int_{\beta_1}^{\beta_2}V(u)d\b} \; \to \; 0
\ee
\qed
\begin{lemma} The following expression holds:
\be\label{expr-delta2}
 \Delta_2 G \; =\; - \beta\, R \left [
\sum_{k=1}^R  <G \, q_{\,l,\,R+1}> - (R+1)  <G \, q_{\,R+1,\,R+2}>
 +  <G>\,<q_{1,2}> 
\right] \; .
\ee
\end{lemma}
{\bf Proof.}
In order to obtain the $\Delta_2G$ we are left with the explicit evaluation
of 
the other term in (\ref{delta2}) which simply gives
\bea\nonumber
\av{\Omega(h(\s^{(l)}))}\,\av{\Omega (G)} & = &
\frac{1}{|\Lambda|}\,
\av{\;
\sum_{\sigma^{(l)}}\;
H_{\Lambda}(\s^{(l)}) \;
p_{\Lambda}\,(1)} \,<G>\\ \nonumber
& = &
\av{\;
\sum_{\sigma^{(l)}}\;\sum_{\gamma}\;
{c}_{\Lambda}(\s^{(l)},\gamma)\;
\frac{dp_{\Lambda}\,(1)}{dH_{\Lambda}(\gamma)}}
\,<G>
\quad
\qquad 
\\
& = &
\label{temp2}
- \beta\, <G>[<q_{1,1}>-<q_{1,2}>]
\eea

\noindent
Inserting the (\ref{line4}) and (\ref{temp2}) in Eq. (\ref{delta2})
we obtain the (\ref{expr-delta2}).
\qed

\vspace{1truecm}
{\bf Acknowledgments}. We thank S. Graffi and F. Guerra for many
interesting
discussions on the subject and for their suggestions. We also thank
A.Bovier, C.Newman,
D.Stein, M.Talagrand and F.L. Toninelli.

\end{document}